\newcommand{\ddt}{\frac{\mathrm d}{\mathrm dt}}

\newcommand{\dds}{\frac{\mathrm d}{\mathrm ds}}
\newcommand{\ddsp}{\tfrac{\mathrm d}{\mathrm ds}}

\newcommand{\Ddss}{\frac{\mathrm D^2}{\mathrm ds^2}}

\newcommand{\R}{\mathds R}
\newcommand{\N}{\mathds N}

\documentclass[10pt,letterpaper]{amsart}

\usepackage{amssymb}

\usepackage{epsfig}
\usepackage{times}
\usepackage{dsfont}
\usepackage{hyperref}

\numberwithin{equation}{section}

\title[A variational approach to homogeneous scalar fields]%
{A variational approach to homogeneous scalar fields in General Relativity}

\author[R.\ Giamb\`o ,\ F.\ Giannoni]{Roberto Giamb\`o, Fabio Giannoni}
\address{Dipartimento di Matematica e Informatica \hfill\break\indent
Universit\`a di Camerino\hfill\break\indent Italy}
\email{roberto.giambo@unicam.it, fabio.giannoni@unicam.it}

\author[G.\ Magli]{Giulio Magli}
\address{Dipartimento di Matematica \hfill\break\indent Politecnico di
Milano \hfill\break\indent Italy} \email{magli@mate.polimi.it}

\begin{document}


\theoremstyle{plain}\newtheorem{teo}{Theorem}[section]
\theoremstyle{plain}\newtheorem{prop}[teo]{Proposition}
\theoremstyle{plain}\newtheorem{lem}[teo]{Lemma}
\theoremstyle{plain}\newtheorem{cor}[teo]{Corollary}
\theoremstyle{definition}\newtheorem{defin}[teo]{Definition}
\theoremstyle{remark}\newtheorem{rem}[teo]{Remark}
\theoremstyle{definition}\newtheorem{example}[teo]{Example}

\theoremstyle{plain}\newtheorem*{prob}{Problem}

\begin{abstract}
A result of existence of homogeneous scalar field solutions between
prescribed
configurations is given, using a modified version \cite{VG2} of
Euler--Maupertuis least action variational principle. Solutions are obtained
as
limit of approximating variational problems, solved using techniques
introduced
by Rabinowitz in \cite{Rabi}.
\end{abstract}

\maketitle

\section{Introduction}\label{sec:intro}
Observational cosmology suggests that our universe has entered a stage of
accelerated expansion (see \cite{riess, tsu} and references therein). The
reason for that is the so called \emph{dark energy}, that constitutes the
most
of the energy density of the whole universe. The existence of a cosmological
constant can be called as a responsible for dark energy, but some physical
problems arise from this interpretation. The most important alternative
physical interpretation for dark energy origin calls into play scalar field
models of spacetime (see \cite{nunes}, and references therein). In the
latest
years of theoretical physics, the belief for existence of \emph{zero--spin}
particles -- whose description is given in terms of a wave \emph{scalar}
function -- has gained supporters, although there is no observational
evidence
yet.

The general scalar field spacetime is a Lorentzian manifold $(M,g)$ such
that
the metric $g$ satisfies the Einstein field equation
\begin{equation}\label{eq:efe}
R-\frac12 S\,g=8\pi\,T.
\end{equation}
On the left hand side, $R$ and $S$ are respectively Ricci tensor field and
scalar curvature function of $g$ (see \cite{ON}), and are completely
determined
by components of $g$ and their partial derivatives up to second order. On
the
righthand side, $T$ is the energy momentum tensor field, that in this case
is
completely determined by a scalar function $\phi$ on $M$, and a potential
function $V(\phi)$. Its expression, evaluated on a couple $(u,v)$ of vectors
belonging to $T_xM$ (the tangent space to $M$ in a point $x\in M$),
reads\footnote{the $4\pi$ factor on the left hand side of \eqref{eq:T} is
just
to simplify the form of Einstein field equations, getting rid of the $\pi$
factor in \eqref{eq:efe1}--\eqref{eq:efe2} below.} \cite{wald}
\begin{equation}\label{eq:T}
4\pi T(x)[u,v]=\mathrm d\phi(x)[u]\cdot\mathrm
d\phi(x)[v]-\left[\frac12g(\nabla\phi(x),\nabla\phi(x))+V(\phi(x))\right]g(u,v),
\end{equation}
where $\nabla\phi$ denotes the Lorentzian gradient vector field of $\phi$.

In this paper we will make the assumptions that the spacetime is
spatially homogeneous with Bianchi I type symmetry \cite{Kra}.
This allows us to choose a convenient coordinate system
$x=(x^0=t,x^1,x^2,x^3)$, such that the metric can be written in
the form
\begin{equation}\label{eq:g}
g=-\mathrm dt\otimes\mathrm dt+a^2(t)\left[\mathrm dx^1\otimes\mathrm
dx^1+\mathrm dx^2\otimes\mathrm dx^2+\mathrm dx^3\otimes\mathrm dx^3\right]
\end{equation}
and the scalar field $\phi$ that  determines $T$ \eqref{eq:T} is a function
$\phi=\phi(t)$ of the variable $t$ only. With the above assumption, a
convenient set for Einstein equations \eqref{eq:efe} is the following:
\begin{subequations}
\begin{align}
&(G^0_0=8\pi T^0_0):\qquad-\frac{3\dot
a^2}{a^2}=-(\dot\phi^2+2V(\phi)),\label{eq:efe1}\\
&(G^1_1=8\pi T^1_1):\qquad-\frac{\dot a^2+2a\ddot
a}{a^2}=(\dot\phi^2-2V(\phi)).\label{eq:efe2}
\end{align}
\end{subequations}
where the dot denotes differentiation with respect to $t$.

We will be interested in the problem of determining solutions of
\eqref{eq:efe1}--\eqref{eq:efe2} with fixed endpoints. The metric becomes
singular when $a(t)$ vanishes in the past or in the future -- corresponding
to
big--bang or big--crash singularity, respectively. In this paper we will
want
to avoid this situation, so we will consider pieces of evolution where
$a(t)$
keeps positive. The central result of the paper is the following theorem:
\begin{teo}\label{thm:main}
Let $a_0,a_1\in\R^+$, $\phi_0,\phi_1\in\R$ be such that
\begin{equation}\label{eq:assum1}
3\min\{a_0,a_1\}(a_1-a_0)^2>\max\{a_0,a_1\}(\phi_1-\phi_0)^2,
\end{equation}
and let $V\in\mathcal{C}^1(\R,\R)$ such that
\begin{equation}\label{eq:assum2}
V(\phi)>0,\qquad\forall\phi\in\R.
\end{equation}
Then, there exists $T>0$ and $(a(t),\phi(t))\in\mathcal{C}^2([0,T],\R^2)$
solutions of \eqref{eq:efe1}--\eqref{eq:efe2} with the boundary conditions
\begin{equation}\label{eq:boundary}
a(0)=a_0,\qquad a(T)=a_1,\qquad\phi(0)=\phi_0,\qquad\phi(T)=\phi_1.
\end{equation}
\end{teo}

Note that the above result cannot be seen as a consequence of
well-known theory on existence of solutions for Cauchy problems in
General Relativity (see pioneering work by Bruhat \cite{BR} and
following literature). Here, indeed, we do not fix both
fundamental forms on a Cauchy surface, and find evolution from it,
but we show that, under condition above, the set of initial data
can be completed in order to reach a certain configuration from a
given one. The solutions under our study stay regular in the
interval $[0,T]$, since $a(t)>0$. Nevertheless, we believe that
this approach may be useful to find existence results for scalar
field solutions evolving to singularity, a topic which is of
relevant interest in the problems related to the Cosmic censorship
conjecture. In fact, although  existence and causal structure of
the singularities in matter-filled spacetimes has been so far
widely investigated in the case of fluid-elastic matter (see
\cite{ns} and references therein) for scalar fields the situation
is fully understood only in the very special case of non
self-interacting, massless particles \cite{ch94,ch99}. Homogeneous
collapse with potential has been treated only in special cases, by
Joshi \emph{et al} (who have also investigated  on loop quantum
gravity effects in \cite{j2}) and in the work \cite{homosf},where
the collapse features are characterized though the dependence of
the energy density on the scale factor $a$; an important open
question in which models with potentials have been used is also
that of cosmic censorship violation in AdS \cite{hor1,hor2}.

We will cast the above problem into a suitable variational framework. The
use
of a variational approach in General Relativity study is not a novelty, of
course: for instance, some of the most important results in relativistic
gravitational lensing problem are reviewed in \cite{Per}. In the present
paper,
actually, the variational approach is used to determine solutions of
Einstein
field equations, i.e. spacetimes. Hilbert--Palatini action \cite{bbb,wald},
indeed, provides a functional whose critical points with respect to
variation
of the metric $g$ yields solutions of \eqref{eq:efe}. The particular case
under
our study produces the functional \eqref{eq:func1}, which is an integral
made
on the interval $[0,T]$ of definition of the solutions. Of course, since $T$
is
let free in principle, it must be treated as an unknown for the system, but
this problem may be circumvented by using the functional \eqref{eq:func2} in
the space of curves reparameterized on the interval $[0,1]$. Although the
functional \eqref{eq:func2} seems in principle more complicated to deal with
than \eqref{eq:func1}, critical point existence can be obtained as a limit
of a
sequence of variational problems, for which Rabinowitz' Saddle Point Theorem
\cite{Rabi} techniques apply.

The outline of the paper is the following. Section \ref{sec:var} is devoted
to
an exposition of the variational formulation for the problem; Section
\ref{sec:theory} briefly outlines the general theory that applies to the
approximating variational problems introduced in Section \ref{sec:approx},
and
studied in Section \ref{sec:pen}. Final section \ref{sec:proof} contains the
proof of the main Theorem \ref{thm:main}.

\section{The variational principle}\label{sec:var}

In this section we collect some basic facts that lead to the variational
formulation of homogeneous scalar field theory. As is well known, solutions
of
Einstein field equations \eqref{eq:efe} are related to critical points of
the
Hilbert--Palatini action functional
\begin{equation}\label{eq:HP1}
\mathcal{I}=\int_M \sqrt{-\det g}\,(L_g+L_f) \,\mathrm dV,
\end{equation}
where $L_g$ and $L_f$ are Lagrangian scalar function related to the
contribution of gravitation (the metric $g$) and the external source (the
energy momentum tensor) respectively. The Lagrangian $L_g$ is given, in
general, by $\tfrac1{16\pi}S$, that in this case reads
\begin{equation}\label{eq:Lg}
L_g=\frac{3}{8\pi}\frac{\dot a^2(t)+a(t)\,\ddot a(t)}{a^2(t)},
\end{equation}
whereas $L_f$ depends on the source of matter, and in this case takes the
form
\begin{equation}\label{eq:Lf}
L_f=-\frac1{4\pi}\left(\frac12 g(\nabla\phi,\nabla\phi)+V(\phi)\right)
=\frac1{8\pi}(\dot\phi^2(t)-2V(\phi(t))).
\end{equation}
Since all the unknown functions depends on $t$ only, we can reduce the
volume
integral in \eqref{eq:HP1} to an integral made on the interval $[0,T]$ of
definition  of $a(t)$ and $\phi(t)$. Using \eqref{eq:Lg} and \eqref{eq:Lf},
therefore, the functional we are dealing with is found to be
\begin{equation}\label{eq:func0}
\mathcal{L}(a,\phi)=\int_0^T 3\big(\dot a^2(t)+a(t)\,\ddot a(t)\big)
a(t)+a^3(t)(\dot\phi^2(t)-2V(\phi(t))\,\mathrm dt.
\end{equation}
We can integrate by parts the term in \eqref{eq:func0} containing $\ddot
a(t)$.
We ignore the contribution of the boundary term $3a^2\dot a\big|_0^T$ coming
from the integration. This exactly amounts \cite{wald} to modify the
functional
\eqref{eq:func0}, adding the contribution $\tfrac1{16\pi}\int_{\partial M}
K$
of the trace of the extrinsic curvature $K$ integrated along the boundary
$\partial M$ of the spacetime.

All in all, we obtain
\begin{equation}\label{eq:func1}
\mathcal{L}(a,\phi)=\int_0^T 3 a(t)\dot a^2(t)-a^3(t)\dot\phi^2(t)+2 a^3(t)
V(\phi(t))\,\mathrm dt,
\end{equation}
where we have also performed an unsubstantial overall change of sign inside
the
integral. The following proposition holds.
\begin{prop}\label{thm:var}
If $(a,\phi)\in\mathcal{C}^2(\R^+,\R)$ solves Euler--Lagrange equation for
$\mathcal{L}$, and
\begin{equation}\label{eq:initcond}
3\dot a(0)^2=a_0^2(\dot\phi(0)^2+2V(\phi_0)),
\end{equation}
then it is a solution for homogeneous scalar field equation
\eqref{eq:efe1}--\eqref{eq:efe2}.
\end{prop}
\begin{proof}
This is a standard result, consequence of N\"{o}ther's Theorem, but an
\emph{ad--hoc} proof can be easily given. Indeed, Euler--Lagrange equations
for
the functional \eqref{eq:func1} read
\begin{subequations}
\begin{align}
&\dot a^2+2a\ddot a=-a^2(\dot\phi^2-2V(\phi)),\label{eq:EL1}\\
&\ddot\phi+V'(\phi)=-3\frac{\dot a}a\dot\phi,\label{eq:EL2}
\end{align}
\end{subequations}
where also the condition $a\ne 0$ has been used. Using these equations, the
quantity $a(3\dot a^2-a^2(\dot\phi^2+2V))$ is easily seen to be constant
w.r.t.
$t$, and then if \eqref{eq:efe1} holds at initial time -- which is just
condition \eqref{eq:initcond} -- then it vanishes during the whole
evolution.
Since $a\ne 0$ we therefore obtain \eqref{eq:efe1}. Equation \eqref{eq:efe2}
is
equivalent to \eqref{eq:EL1}.
\end{proof}

\begin{rem}
A \emph{partial} converse of Proposition above can be of course given, using
the following identity:
\begin{equation}\label{eq:Bianchi}
T^\mu_{\,\,0;\mu}=-2\dot\phi\left(\ddot\phi+V'(\phi)+3\frac{\dot
a}a\dot\phi\right).
\end{equation}
The left hand side above is the 0-component of the divergence of $T$
\cite{wald}, and it vanishes if Einstein equations holds, due to
conservation
of energy (Bianchi identity). Therefore, if $\dot\phi$ is everywhere
nonzero,
\eqref{eq:EL2} holds. Equation \eqref{eq:EL2} is also known in literature as
Klein--Gordon equation, and its intrinsic expression reads
$\square\phi=V'(\phi)$, where
$\square\phi=g^{\alpha\beta}\nabla_a\nabla_\beta\phi$ is the d'Alembert
operator with respect to the Lorentzian metric $g$.
\end{rem}

Actually, the converse result can be improved further:

\begin{prop}
Se $(a,\phi):[0,T]\to\R^+\times\R$ are $\mathcal C^2$ solutions of
\eqref{eq:efe1}--\eqref{eq:efe2} with $\phi\not\equiv\phi_0$ on $[0,T]$,
then
solve \eqref{eq:EL1}--\eqref{eq:EL2}. In particular \eqref{eq:initcond}
holds.
\end{prop}

\begin{proof}
Only \eqref{eq:EL2} must be proved. Let $G(t)$ be the (continuous) function
$G(t):=\ddot\phi(t)+V'(\phi(t))+3\frac{\dot a(t)}{a(t)}\dot\phi(t)$, and
assume
by contradiction the existence of $t_0\in[0,T]$ such that $G(t_0)\ne 0$.
Therefore, considered the (closed) set
$C=\{t\in[0,T]\,:\,\dot\phi(t)=0\}\subsetneq [0,T],$ Bianchi identity
\eqref{eq:Bianchi} implies the existence of a closed interval
$[\alpha,\beta]$ such that
\begin{equation}\label{eq:max}
t_0\subset[\alpha,\beta]\subseteq C,
\end{equation}
and that is maximal with respect to this property \eqref{eq:max}. We will
show
$G=0$ on $[\alpha,\beta]$, getting a contradiction.

Since $[\alpha,\beta]$ is strictly contained in $[0,T]$ then $\alpha>0$ or
$\beta<T$ (or both). Let us assume  for instance $\alpha>0$. Since
$\dot\phi_{\vert[\alpha,\beta]}=0$, and $\phi$ is $\mathcal C^2$, then
$\ddot\phi_{\vert[\alpha,\beta]}=0$ and $\phi(t)=\phi(\alpha),\forall
t\in[\alpha,\beta]$. These facts imply
\begin{equation}\label{eq:1}
G(t)=V'(\phi(\alpha)),\qquad\forall t \in[\alpha,\beta],
\end{equation}
Maximality of $[\alpha,\beta]$ w.r.t. \eqref{eq:max} implies the existence
of a
sequence  $\alpha_n\not\in C$, $\alpha_n\to\alpha^-$. Then
$\dot\phi(\alpha_n)\ne 0$ and therefore $G(\alpha_n)=0$ by
\eqref{eq:Bianchi},
therefore continuity of $G$ and \eqref{eq:1} finally imply
\[
0=\lim_{n\to\infty}G(\alpha_n)=\lim_{t\to\alpha^-}G(t)=V'(\phi(\alpha)),
\]
obtaining $G(t)=0$ on $[\alpha,\beta]$.
\end{proof}

\begin{rem}\label{rem:elas}
If $\phi$ is everywhere constant one can easily find counterexamples where
\eqref{eq:efe1}--\eqref{eq:efe2} hold but \eqref{eq:EL2} do not. This is a
well
known fact in relativistic elasticity theory, where equivalence between
Bianchi
identity and Euler--Lagrange equations (i.e. N\"{o}ther theorem) holds under
the assumption that the deformation tensor has maximum rank. We refer the
reader to \cite{KM} for further details on this topic. In scalar field
theory,
the gradient $\nabla\phi$ plays the role of this deformation tensor. For our
purposes here, anyway, we will only use Proposition \ref{thm:var}.
\end{rem}

The problem of finding solutions of \eqref{eq:efe1}--\eqref{eq:efe2} with
fixed
endpoints is brought to the study of critical points of the
functional \eqref{eq:func1} with fixed endpoints and under the initial
condition \eqref{eq:initcond}. Of course, since the arrival time $T$ is left
free so far, one can reparameterize the functions on the interval $[0,1]$,
with
the obvious drawback to promote $T$ as a new unknown of the problem. But
this
problem can be overcome, applying the following (general) variational
principle. \cite{Ambr,VG}

\begin{teo}\label{thm:princ}
Let $(\mathfrak{M},\mathfrak{g})$ be a semi--Riemannian manifold, $W$ a
$\mathcal{C}^1$ function on $\mathfrak{M}$, $E\in\R$ and
$p,q\in\mathfrak{M}$.

\begin{enumerate}
\item If $y:[0,1]\to\mathfrak{M}$ is a critical point for the functional
\begin{equation}\label{eq:F2}
{\mathfrak{F}}(y)=\left(\int_0^1\frac12\mathfrak{g}\big(\dds y(s),\dds
y(s)\big)\,\mathrm ds\right)\cdot\left(\int_0^1E-W(y(s))\,\mathrm ds\right)
\end{equation}
with positive critical value, in the space of $\,\,\mathcal{C}^2$ curves
defined in $[0,1]$, such that $y(0)=p$, $y(1)=q$, called
\begin{equation}\label{eq:T0}
T_0=\left(\frac{\int_0^1\frac12\mathfrak{g}\big(\dds y,\dds y\big)\,\mathrm
ds}{\int_0^1E-W(y(s))\,\mathrm ds}\right)^{1/2},
\end{equation}
then the curve $x:[0,T_0]\to\mathfrak{M}$, $x(t)=y(T_0 s)$ is a critical
point
for the functional
\begin{equation}\label{eq:F1}
\mathfrak{L}(x)=\int_0^T\frac12\mathfrak{g}\big(\ddt x(t),\ddt x(t)\big)
-W(x(t))\,\mathrm dt,
\end{equation}
with $T=T_0$, in the space of $\mathcal{C}^2$ curves satisfying the
conditions
\begin{equation}\label{eq:ic}
x(0)=p,\qquad x(T)=q,\qquad \frac12\mathfrak{g}\big(\ddt x(t),\ddt
x(t)\big)+W(x(t))=E.
\end{equation}
\item
Viceversa, let us fix $T>0$, and let $x:[0,T]\to\mathfrak{M}$ be a critical
point for the functional \eqref{eq:F1} in the space of $\,\,\mathcal{C}^2$
curves $\gamma:[0,T]\to\R$ satisfying conditions \eqref{eq:ic}. If
$\int_0^T\frac12\mathfrak{g}(\ddt x(t),\ddt x(t))\,\mathrm dt\ne0$, then the
reparameterization $y:[0,1]\to\mathfrak{M}$ of $x$ on $[0,1]$, (i.e.
$y(s)=x(T\,s)$), is a critical point for the functional \eqref{eq:F2}, with
positive critical value, in the space of $\,\,\mathcal{C}^2$ curves defined
in
$[0,1]$, such that $y(0)=p$, $y(1)=q$.
\end{enumerate}
\end{teo}
\begin{proof}
Let $y(s):[0,1]\to\mathfrak{M}$ be a critical point for ${\mathfrak{F}}$
\eqref{eq:F2}. Fixed endpoint first variation of this functional reads
\begin{multline*}
\mathrm d{\mathfrak{F}}(y)[\eta]=\\
\int_0^1 E-W(y)\,\mathrm ds \,\int_0^1\mathfrak{g}\big(\dds
y,\dds\eta\big)\,\mathrm ds-\int_0^1\frac12\mathfrak{g}\big(\dds y,\dds
y\big)\,\mathrm ds\,\int_0^1W'(y)[\eta]\,\mathrm ds,
\end{multline*}
and integrating by part we get the following equation
\begin{equation}\label{eq:var1}
\left(\int_0^1 E-W(y(s))\,\mathrm ds\right)\Ddss
y(s)+\left(\int_0^1\frac12\mathfrak{g}\big(\dds y(s),\dds y(s)\big)\,\mathrm
ds\right)\nabla W(y(s))=0
\end{equation}
that is, using the value of $T_0$ given by \eqref{eq:T0} -- well defined
since
the critical value is positive
\begin{equation}\label{eq:Lagr2}
\frac1{T_0^2}\Ddss y(s)+\nabla W(y(s))=0,\qquad\forall s\in [0,1].
\end{equation}
Let $x:[0,T_0]\to\mathfrak{M}$ be the reparameterization of $y$ on the
interval
$[0,T_0]$ (i.e. $x(t)=y(t/T_0)$). Therefore, equation \eqref{eq:Lagr2}
becomes
nothing but Euler--Lagrange equation for the functional \eqref{eq:F1} with
$T=T_0$. Moreover, contracting the left hand side of \eqref{eq:Lagr2} with
$\ddsp y(s)$, we obtain the existence of a constant $K$ such that
\begin{equation}\label{eq:cons-rep}
\frac1{T_0^2}\frac12\mathfrak{g}\big(\dds y(s),\dds
y(s)\big)+W(y(s))=K,\qquad\forall s\in[0,1].
\end{equation}
Integrating both sides above between 0 and 1, and using \eqref{eq:T0}, we
obtain $E=K$, and so
\[
\frac12\mathfrak{g}\big(\ddt x(t),\ddt x(t)\big)+W(x(t))=E,\qquad\forall
t\in[0,T_0].
\]

Conversely, let $x(t):[0,T]\to\mathfrak{M}$ be a critical point for
\eqref{eq:F1} in the space of $\,\,\mathcal{C}^2$ curves defined in $[0,T]$,
satisfying \eqref{eq:ic},  and $y(s):[0,1]\to\mathfrak{M}$ be its
reparameterization: $y(s)=x(T\,s)$. Since $\dds=T\ddt$, \eqref{eq:cons-rep}
holds with $T_0$ and $K$ replaced by $T$ and $E$ respectively. Integrating
both
sides of \eqref{eq:cons-rep} in $[0,1]$ implies that $T$ is equal to the
value
of $T_0$ given by \eqref{eq:T0}. Moreover, $\dds=T\ddt$ also implies
\eqref{eq:Lagr2}, and substituting the value of $T_0$ given by \eqref{eq:T0}
we
find that $y(s)$ satisfies \eqref{eq:var1}, that is critical point equation
for
the functional \eqref{eq:F2}, and the proof is complete.
\end{proof}

\begin{rem}
The variational principle given above, is actually a sort of modified
version
of the classical Euler--Maupertuis least action principle. The above proof
is
an adaptation of the one given in \cite{VG} for the Euclidean case, and may
not
completely stress the link with the classical principle. For a deeper
insight,
we refer the reader to the review \cite{VG2} by the same author.
\end{rem}

Applying this variational principle to functional \eqref{eq:func1} (with
$E=0$), the following problem provides solutions of homogeneous scalar field
equation with fixed endpoints:
\begin{prob}
Let $a_0,a_1\in\R^+$, $\phi_0,\phi_1\in\R$, and $V\in\mathcal{C}^1(\R,\R)$.

Find the critical points of the functional
\begin{equation}\label{eq:func2}
F(a,\phi)=\left(\int_0^1 3 a(t)\dot a^2(t)-a^3(t)\dot\phi^2(t)\,\mathrm
dt\right)\cdot\left(\int_0^1 2 a^3(t) V(\phi(t))\,\mathrm dt\right),
\end{equation}
with positive critical value, in the space of $\,\,\mathcal{C}^2$ curves
$(a,\phi):[0,1]\to\R^+\times\R$ such that
\begin{equation}\label{eq:bound}
a(0)=a_0,\qquad a(1)=a_1,\qquad\phi(0)=\phi_0,\qquad\phi(1)=\phi_1.
\end{equation}
\end{prob}

\section{The functional framework and the abstract critical points
theory}\label{sec:theory}

We first recall the classical notion of Palais--Smale condition.
\begin{defin}\label{def:PS}
Let $\mathfrak{X}$ a Hilbert manifold of class $\mathcal{C}^1$ and
$f\in\mathcal{C}^1(\mathfrak{X},\R)$. We say that $f$ satisfies the
\emph{Palais--Smale condition at level} $c$ (abbrev. $(PS)_c$) if any
sequence
$\{x_n\}_{n\in\N}\subset\mathfrak{X}$ such that
$$
f(x_n)\to c,\qquad\text{and\ }\nabla f(x_n)\to 0
$$
(where $\nabla f$ represents the gradient of $f$ w.r.t. the Hilbert
structure
of $\mathfrak{X}$) has a converging subsequence in $\mathfrak{X}$. The
sequence
$\{x_n\}$ with properties above is called a \emph{Palais--Smale sequence}
for
$f$.
\end{defin}

\begin{defin}\label{def:cp}
We say that $x\in\mathfrak{X}$ is a \emph{critical point} of $f$ if $\nabla
f(x)$=0. A value $c\in\R$ such that there exists a critical point $x$ with
$f(x)=c$ is a \emph{critical value} for $f$. A value $c\in\R$ which is not
critical will be called \emph{regular}.
\end{defin}

The following lemma is a slight modification of a well known deformation
lemma
(see e.g. \cite{Palais}) and it will be used in Theorem \ref{thm:teo3.4}
later.

For any $d\in\R$ set $f^d=\{x\in\mathfrak{X}\,:\,f(x)\le d\}$.

\begin{lem}\label{thm:lem3.3}
Let $\mathfrak{X}$ be a Hilbert manifold and
$f\in\mathcal{C}^1(\mathfrak{X},\R)$. Let $a<b\in\R$. Assume that
\begin{enumerate}
\item\label{itm:ass1} $f$ satisfies $(PS)_c$, $\forall c\in[a,b]$;
\item\label{itm:ass2} the strip $\{x\in\mathfrak{X}\,:\,a\le f(x)\le
b\}\subset\mathfrak{X}$ is complete
(w.r.t. the Hilbert structure of $\mathfrak{X}$);
\item\label{itm:ass3} each value $c\in[a,b]$ is regular for $f$.
\end{enumerate}
Then, there exists a homotopy $\mathfrak{h}:[0,1]\times f^b\to f^b$ such
that
\begin{itemize}
\item $\mathfrak{h}(0,x)=x,\qquad\forall x\in f^b$;
\item $h(\tau,x)=x,\qquad\forall x\in f^a,\forall\tau\in[0,1]$;
\item $h(1,x)\in f^a,\qquad\forall x\in f^b$.
\end{itemize}
\end{lem}
If the gradient of $f$ is locally Lipschitz continuous, the above Lemma can
be
proved using the gradient flow of $f$: indeed, if $f$ satisfies $(PS)_c$
$\forall c\in[a,b]$, then $\Vert\nabla f(x)\Vert_\mathfrak{X}$ is uniformly
bounded away from 0 when $x$ satisfies $a\le f(x)\le b$. Thanks also to
assumption \eqref{itm:ass2}, the solutions of
\[
\left\{
\begin{aligned}
&\ddt\eta(\tau)=\nabla f(\eta(\tau)),\\
&\eta(0)=x,
\end{aligned}
\right.
\]
with $a\le f(x)\le b$, allow to send $f^b$ to $f^a$ in a finite time. If
$\nabla f$ is only continuous, we can use the so--called \emph{pseudo
gradient}
vector field, first introduced by Palais \cite{Palais}, which is locally
Lipschitz continuous. Also in this case the sublevel $f^b$ can be deformed
on
$f^a$, provided the assumptions \eqref{itm:ass1} and \eqref{itm:ass2} hold.

Using Lemma \ref{thm:lem3.3}, the following result, that is a slight
modification of the well known Rabinowitz' Saddle Point Theorem, \cite{Rabi}
holds.

\begin{teo}\label{thm:teo3.4}
Let $\mathfrak{X}=\Omega\times{Y}$, where $\Omega$ is a Hilbert manifold and
${Y}$ is a finite dimensional affine space. Let $\Vert\cdot\Vert$ denote the
norm on ${Y}$, and let $f\in\mathcal{C}^1(\mathfrak{X},\R)$. Assume that
\begin{enumerate}
\item\label{itm:assi}
there exists $\omega_0\in\Omega$, $e_0\in {Y}$, and $R>0$ such that, called
$B_R(e_0)=\{e\in E\,:\,\Vert e-e_0\Vert\le R\}$, it is
\[
b_0\equiv\sup_{e\in\partial
B_R(e_0)}f(\omega_0,e)<b_1\equiv\inf_{\omega\in\Omega}f(\omega,e_0);
\]
\item\label{itm:assiii}
if $b_2=\sup_{e\in B_r(e_0)}f(\omega_0,e)$, the strip
$\{x\in\mathfrak{X}\,:\,b_1\le f(x)\le b_2\}\subset\mathfrak{X}$ is
complete;
\item\label{itm:assii}
$f$ satisfies $(PS)_c$ at any $c\in[b_1,b_2]$.
\end{enumerate}
Then, there exists a critical value $c$ for $f$ in $[b_1,b_2]$.
\end{teo}

The proof can be obtained adapting the scheme developed in \cite{Rabi}.
However, the idea  behind is quite simple. If, by contradiction, $[b_1,b_2]$
is
made by regular value only, by Lemma \ref{thm:lem3.3} there xists a homotopy
sending $f^{b_2}$ to $f^{b_1}$, letting $f^{b_1}$ fixed. Therefore, using
the
projection on ${Y}$ and the retraction of ${Y}$ on $B_R(e_0)$, and observing
that $b_0<b_1$, one can define a homotopy that sends $B_R(e_0)$ to its
boundary
$\partial B_R(e_0)$ -- recall that $b_2=\sup_{e\in B_R(e_0)}f(x_0,e)$ -- and
that lets $\partial B_R(e_0)$ fixed, which is impossible in finite
dimension.

\section{An approximation scheme}\label{sec:approx}
By assumption \eqref{eq:assum1} of Theorem \ref{thm:main}, we know that
$a_0\ne
a_1$. To fix ideas, without loss of generality we can assume
\begin{equation}\label{eq:4.1}
a_0>a_1,
\end{equation}
so that \eqref{eq:assum1} becomes
\[
3 a_1(a_1-a_0)^2>a_0^3 (\phi_1-\phi_0)^2.
\]
We can choose constants $m,M$ such that $0<m<a_1<a_0<M<+\infty$ and
\begin{equation}\label{eq:4.2}
3m(a_1-a_0)^2-M^3(\phi_1-\phi_0)^2>0.
\end{equation}
Consider the Hilbert manifold
\begin{equation}\label{eq:4.3}
\Omega=\{a\in H^1([0,1],]m,M[)\,:\,a(0)=a_0,\,\,a(1)=a_1\},
\end{equation}
where $H^1([0,1],]m,M[)$ is the set of absolutely continuous functions
defined
on $[0,1]$, with values on $]m,M[$, such that $\int_0^1\dot a^2\,\mathrm
dt<+\infty$. Let us observe that $\Omega$ is a \emph{not} complete Hilbert
manifold with Hilbert structure
\[
\big\langle a_1,a_2\rangle=\int_0^1\dot a_1(t)\dot a_2(t)\,\mathrm dt.
\]
We denote by $\Vert a\Vert_\Omega$ the norm induced by the above inner
product:
\begin{equation}\label{eq:4.4}
\Vert a\Vert_\Omega=\left(\int_0^1\dot a(t)\,\mathrm dt\right)^{1/2}.
\end{equation}
Now set $\phi_*(t)=(1-t)\phi_0+t\phi_1$. Since $V>0$, we have
\begin{equation}\label{eq:4.5}
\inf_{t\in[0,1]}V(\phi_*(t))\equiv v_*>0.
\end{equation}
Let us also consider the affine space
\begin{equation}\label{eq:4.6}
{Y}=\{\phi=\widehat{\phi}+\phi_*\,:\,\widehat{\phi}\in H_0^1([0,1],\R)\},
\end{equation}
where $H^1_0([0,1],\R)=\{\phi\in H^1([0,1],\R)\,:\,\phi(0)=\phi(1)=0\}.$
${Y}$
is a closed affine subspace of $H^1([0,1],\R)$, with norm
\begin{equation}\label{eq:4.7}
\Vert\phi\Vert_{Y}=\left(\int_0^1\dot\phi^2(t)\,\mathrm dt\right)^{1/2}.
\end{equation}
Since $\dim {Y}=+\infty$ we cannot apply Theorem \ref{thm:teo3.4} \emph{as
is}
to our setting, and then we approximate ${Y}$ by a sequence ${Y}_k$, defined
as
follows: for any $k\in\N$ set
\begin{equation}\label{eq:4.7bis}
W_k=\text{span}\{\sin(\pi\,\ell\,t)\,:\,t\in[0,1],\,\ell=1,\ldots,k\},
\end{equation}
and
\[
{Y}_k=\{\phi=\widehat{\phi_k}+\phi_*\,:\,\widehat{\phi_k}\in W_k\}.
\]

\begin{rem}\label{rem:4.1}
Since $\{\sqrt2\sin(\pi\ell\,t)\}_{\ell\in\N}$ is a complete orthonormal
system
of $H^1_0([0,1],\R)$, if $\phi=\widehat{\phi}+\phi_*\in {Y}$ and
$\widehat{\phi_k}$ denotes the projection of $\widehat{\phi}$ on $W_k$, then
$\widehat{\phi_k}\to\widehat{\phi}$ in $H^1_0$, w.r.t. the norm defined in
\eqref{eq:4.7}.
\end{rem}

We shall apply Theorem \ref{thm:teo3.4} to the space
$\mathfrak{X}_k=\Omega\times {Y}_k$. Since $\Omega$ is not complete, and $V$
is
not bounded in general, we modify the functional $F$ \eqref{eq:func2} and
look
for critical points $x_{\epsilon,\lambda}$, with positive critical value, of
a
suitable functional $F_{\epsilon,\lambda}$. Some estimates for the critical
points $x_{\epsilon,\lambda}$ will show that they are critical points for
$F$,
whenever $\epsilon$ is sufficiently small, and $\lambda$ sufficiently large.

Let $\chi:\R^+\to\R^+$ of class $\mathcal{C}^1$, such that $\chi(s)=0$ if
$s\le
0$ and $\chi(s)=s^2$ if $s>0$. Fix $\epsilon\in]0,1]$, and define
\begin{equation}\label{eq:4.8}
U_\epsilon(a)=\chi\left(\frac1{a-m}-\frac1\epsilon\right)+\chi\left(\frac1{M-a}-\frac1\epsilon\right).\
\end{equation}
Moreover, consider $\psi:\R\to\R$ of class $\mathcal{C}^1$ such that
$\psi(s)=s$ if $s\le 0$, $\psi(s)=1$ if $s\ge 1$, and $\psi$ is strictly
increasing on the interval $]0,1[$. Fix $\lambda>0$ and define
\begin{equation}\label{eq:4.9}
V_\lambda(\phi)=\psi\left(V(\phi)-\lambda\right)+\lambda.
\end{equation}
Observe that $V_\lambda=V$ whenever $V(\phi)\le\lambda$. Finally, define
\begin{equation}\label{eq:4.10}
F_{\epsilon,\lambda}(a,\phi)=\int_0^1\left[\left(3a+U_\epsilon(a)\right)\dot
a^2-a^3\dot\phi^2\right]\,\mathrm dt\cdot\int_0^12 a^3
V_\lambda(\phi)\,\mathrm
dt.
\end{equation}
It is a straightforward computation to show that the above functional is
$\mathcal{C}^1$. In the next section we shall show how to apply Theorem
\ref{thm:teo3.4} to $F_{\epsilon,\lambda}$ on the space $\Omega\times
{Y}_k$.

\section{Critical points for the functional
$F_{\epsilon,\lambda}$}\label{sec:pen}

The aim of this section is to prove the following result.

\begin{prop}\label{thm:prop5.5}
For any $k\in\N$ there exists $x_k=(a_k,\phi_k)$ critical point of
$F_{\epsilon,\lambda}$ on $\mathfrak{X}_k$ such that
\[
F_{\epsilon,\lambda}(x_k)\in[b_1,b_2],
\]
where  $b_1,b_2$ are positive, and independent of $k$.
\end{prop}

We  must first show that hypotheses \eqref{itm:assi}--\eqref{itm:assiii} of
Theorem \ref{thm:teo3.4} hold for $F_{\epsilon,\lambda}$ on
$\mathfrak{X}_k=\Omega\times{Y}_k$. The key point will be to show
\eqref{itm:assii}, namely Palais--Smale condition (Definition \ref{def:PS}).
This will be done in Lemma \ref{thm:lem5.4}. First, let us prove the
validity
of hypotheses \eqref{itm:assiii}.

\begin{lem}\label{thm:lem5.3}
Denoted by $F^c$ the sublevel $F^c=\{x\in\Omega\times E\,:\,F(x)\le c\}$,
then
the set $F^c\cap\mathfrak{X}_k$ is complete in $\mathfrak{X}_k$.
\end{lem}

\begin{proof}
Take a Cauchy sequence $x_n=(a_n,\phi_n)\in\Omega\times E_k$. Since the
closure
of $\Omega$ is complete, and so is $E_k$, then there exists
$(a,\phi)\in\overline{\Omega}\times E_k$ such that $a_n\to a$ in $H^1$ and
$\phi_n\to\phi$ in $H^1$. Now $\int_0^1 a_n^3 V_\lambda(\phi_n)\,\mathrm
dt\to\int_0^1a^3V_\lambda(\phi)\,\mathrm dt$ which is strictly  positive,
and
$\int_0^1\left(3 a_n\dot a_n^2-a_n^3\dot\phi_n^2\right)\,\mathrm
dt\to\int_0^13
a\dot a^2-a^3\dot\phi^2\,\mathrm dt$.

If $a(t)\in]m,M[\,\forall t$, then $(a_n,\phi_n)$ converges in $\Omega\times
E_k$. The proof is complete as one shows that there is no possibility that
some
$\overline{t}\in]0,1[$ exists such that either $a(\overline{t})=m$ or
$a(\overline{t})=M$. By contradiction, suppose for instance
\begin{equation}\label{eq:5.5}
\exists \overline{t}\,:\,a(\overline{t})=m,\quad a(t)>m\,\,\forall
t\in[0,\overline{t}[.
\end{equation}
Observe that $F_{\epsilon,\lambda}(a_n,\phi_n)\le c$, $\int_0^1a^3
V_\lambda(\phi)\,\mathrm dt>0$, and $\int_0^1\left(3 a \dot
a^2-a^3\dot\phi^2\right)\,\mathrm dt$ is finite. We will show that the
hypothesis \eqref{eq:5.5} implies
\[
\int_0^{\overline{t}} U_\epsilon(a_n)\dot a_n^2\,\mathrm dt\to+\infty,
\]
obtaining a contradiction.

Since $a_n\to a$ uniformly, $\forall\epsilon>0$ there exists
$\overline{s}<\overline{t}$ and $n_0\in\N$ such that
\[
|a_n(t)-m|\le\epsilon,\qquad\forall
t\in[\overline{s},\overline{t}],\,\forall
n\ge n_0.
\]
Fix $\epsilon$ such that $m+\epsilon<M$. Then, recalling the definition of
$U_\epsilon$, it will suffice to show that
\begin{equation}\label{eq:5.6}
\lim_{n\to\infty}\int_{\overline{s}}^{\overline{t}}\frac{\dot
a_n^2}{(a_n-m)^2}\,\mathrm dt=+\infty,
\end{equation}
which easily follows from the following estimate:
\begin{multline*}
\left(\int_{\overline{s}}^{\overline{t}}\frac{\dot
a_n^2}{(a_n-m)^2}\,\mathrm
dt\right)^{1/2}\ge\frac1{\sqrt{\overline{t}-\overline{s}}}\int_{\overline{s}}^{\overline{t}}\frac{|\dot
a_n|}{a_n-m}\,\mathrm dt\ge\\
\frac1{\sqrt{\overline{t}-\overline{s}}}\left\vert\int_{\overline{s}}^{\overline{t}}\frac{\dot
a_n}{a_n-m}\,\mathrm dt\right\vert=\frac1{\sqrt{\overline{t}-\overline{s}}}
\left\vert \log\left( \frac{a_n(\overline{t})-m}{a_n(\overline{s})-m}\right)
\right\vert,
\end{multline*}
that diverges because $a_n(\overline{t})\to a(\overline{t})=m$ whereas
$a_n(\overline{s})\to a(\overline{s})<m$.
\end{proof}

The lemma below show that Palais--Smale condition actually holds in every
closed interval of $\R$.

\begin{lem}\label{thm:lem5.4}
Fixed two values $c_1,c_2$ such that $0<c_1<c_2<+\infty$, the functional
$F_{\epsilon,\lambda}$ satisfies $(PS)_c$, $\forall c\in[c_1,c_2]$.
\end{lem}

\begin{proof}
Let $(a_n,\phi_n)$ be a Palais--Smale sequence for $F_{\epsilon,\lambda}$
such
that $F_{\epsilon,\lambda}(a_n,\phi_n)\in[c_1,c_2],\,\forall n\in\N$. Since
$\nabla F_{\epsilon,\lambda}(a_n,\phi_n)$ is infinitesimal we have,
$\forall\theta\in W_k$ (recall \eqref{eq:4.7bis})
\[
\vert\nabla
F_{\epsilon,\lambda}(a_n,\phi_n)[0,\theta]\vert\le\delta_n\Vert\theta\Vert_{H^1_0},
\qquad\text{with\ }\delta_n\to 0.
\]
Therefore
\begin{multline*}
\left\vert\left(\int_0^1 2a_n^3 V_\lambda(\phi_n)\,\mathrm
dt\right)\int_0^1\left(-2a_n^3\dot\phi_n\dot\theta\right)\,\mathrm
dt+\right.\\
\left.\left(\int_0^1\left(3a_n+U_\epsilon(a_n)\right)\dot
a_n^2-a_n^3\dot\phi_n^2\,\mathrm dt\right)\int_0^1 2
a_n^3V'_\lambda(\phi_n)[\theta]\,\mathrm
dt\right\vert\le\delta_n\Vert\theta\Vert_{H^1_0}.
\end{multline*}
Multiplying end terms of the inequality chain above by the bounded and
strictly
positive quantity $\int_0^12 a_n^3V_\lambda(\phi_n)\,\mathrm dt$ we have
\begin{multline}\label{eq:5.7}
\left|\left(\int_0^12a_n^3V_\lambda(\phi_n)\,\mathrm
dt\right)^2\int_0^1\left(-2 a_n^3\dot\phi_n\dot\theta\right)\,\mathrm
dt+\right.\\
\left.F_{\epsilon,\lambda}(a_n,\phi_n)\int_0^1
2a_n^3V'_\lambda(\phi_n)\theta\,\mathrm
dt\right\vert\le\widehat{\delta_n}\Vert\theta\Vert_{H^1_0},
\end{multline}
where $\widehat{\delta_n}:=\delta_n\int_0^12 a_n^3V_\lambda(\phi_n)\,\mathrm
dt\to 0$. Since $\phi_n-\phi_*\in W_k$ we can choose $\theta=\phi_n-\phi_*$,
and observe the following facts:
\begin{itemize}
\item $\int_0^12a_n^3 V_\lambda(\phi_n)\,\mathrm dt$ is bounded away from 0,
independently of $n$,
\item $F_{\epsilon,\lambda}(a_n,\phi_n)$ is bounded,
\item $a_n^3 Vì_\lambda(\phi_n)$ is uniformly bounded, and
\item
$\Vert\phi_n-\phi_*\Vert_{L^\infty}\le\Vert\dot\phi-(\phi_1-\phi_0)\Vert_{L^1}
\le\Vert\dot\phi-(\phi_1-\phi_0)\Vert_{L^2}$.
\end{itemize}
But $\int_0^1a_n^3\dot\phi_n\dot\theta\,\mathrm dt=\int_0^1
a_n^3\left(\dot\phi_n^2-\dot\phi_n(\phi_1-\phi_0)\right)\,\mathrm dt$ which
behaves like $\int_0^1\dot\phi_n^2\,\mathrm dt$ if
$\int_0^1\dot\phi_n^2\,\mathrm dt$ diverges -- recall that
$\Vert\theta\Vert_{H^1_0}=\Vert\dot\theta\Vert_{L^2}$. Then, by
\eqref{eq:5.7}
we deduce the existence of a positive constant $D_0$ such that
\begin{equation}\label{eq:5.8}
\int_0^1\dot\phi_n^2\,\mathrm dt\le D_0,\qquad\forall n\in\N.
\end{equation}
Since $m\le a_n\le M$, $F_{\epsilon,\delta}(a_n,\phi_n)\le c$, and $\int_0^1
2a_n^3 V_\lambda(\phi_n)\,\mathrm dt$ is bounded away from zero we deduce,
by
\eqref{eq:5.8}, that $\int_0^1\left(3 a_n+U_\epsilon(a_n)\right)\dot
a_n^2\,\mathrm dt$ is bounded; moreover, since $3 a_n+U_\epsilon(a_n)\ge 3
a_1>0\,\forall n\in\N$, there exists a positive constant $D_1$ such that
\begin{equation}\label{eq:5.9}
\int_0^1\dot a_n^2\mathrm dt\le D_1,\qquad\forall n\in\N.
\end{equation}
By \eqref{eq:5.8} and \eqref{eq:5.9}, there exists $a$ and $\phi$ of class
$H^1$ such that, up to subsequences,
\[
a_n\rightharpoonup a,\qquad\phi_n\rightharpoonup \phi
\]
weakly in $H^1$ and uniformly. Since $F(a_n,\phi_n)$ is bounded from above,
arguing as in Lemma \ref{thm:lem5.3} one proves that $a(t)\in]m,M[,\,\forall
t\in[0,1]$. Therefore it remains to show
\begin{equation}\label{eq:5.10}
\dot a_n\to\dot a\,\qquad\text{in\ }L^2([0,1],\R),
\end{equation}
and
\begin{equation}\label{eq:5.11}
\dot\phi_n\to\dot\phi\,\qquad\text{in\ }L^2([0,1],\R).
\end{equation}
Observing that $\phi_n-\phi\in W_k$, we can choose $\theta=\phi_n-\phi$ in
\eqref{eq:5.7}. Now $F_{\epsilon,\lambda}(a_n,\phi_n)$ is bounded and
\[
\int_0^1 2a_n^3V'_\lambda(\phi_n)[\phi_n-\phi]\,\mathrm dt\to 0
\]
because $2 a_n^3 V'_\lambda(\phi_n)$ is uniformly bounded and
$\sup_{t\in[0,1]}\vert\phi_n(t)-\phi(t)\Vert\to 0$, while
$\Vert\phi_n-\phi\Vert_{H^1_0}$ is bounded and $\int_0^12 a_n^3
V_\lambda(\phi_n)\,\mathrm dt$ is bounded away from zero. Then, by
\eqref{eq:5.7} we deduce
\[
\int_0^1 a_n^3\dot\phi_n(\dot\phi_n-\dot\phi)\,\mathrm dt\to 0.
\]
Moreover, $\int_0^1 a^3\dot\phi_n(\dot\phi_n-\dot\phi)\,\mathrm dt-\int_0^1
a_n^3\dot\phi_n(\dot\phi_n-\dot\phi)\,\mathrm dt\to 0$, since
$\int_0^1\vert\dot\phi_n(\dot\phi_n-\dot\phi)\vert\,\mathrm dt$ is bounded
and
$a_n^3\to a^3$ uniformly. Then
\begin{equation}\label{eq:5.12}
\int_0^1a^3\dot\phi_n(\dot\phi_n-\dot\phi)\,\mathrm dt\to 0,
\end{equation}
and since $\dot\phi_n\rightharpoonup\dot\phi$ weakly in $L^2$ we have
\begin{equation}\label{eq:5.13}
\int_0^1 a^3\dot\phi(\dot\phi_n-\phi)\,\mathrm dt\to 0.
\end{equation}
Therefore, \eqref{eq:5.12}--\eqref{eq:5.13} together gives
\[
\int_0^1 a^3(\dot\phi_n-\dot\phi)^2\,\mathrm dt\to 0,
\]
and since $a^3\ge a_1^3>0,\,\forall t\in[0,1]$ we obtain \eqref{eq:5.11}.

In order to prove \eqref{eq:5.10} note that there exist an infinitesimal
sequence $\epsilon_n$
\[
\left\vert \nabla
F_{\epsilon,\lambda}[\alpha,0]\right\vert\le\epsilon_n\Vert\alpha\Vert_{H^1_0},\qquad\forall\alpha\in
H^1_0,
\]
and then
\begin{multline*}
\left\vert \int_0^1 \left[(3\alpha+U'_\epsilon(a_n)\alpha)\dot
a_n^2+2(3a_n+U_\epsilon(a_n))\dot
a_n\dot\alpha-3a_n^2\alpha\dot\phi_n^2\right]\,\mathrm dt\int_0^1(2 a_n^3
V_\lambda(\phi_n))\,\mathrm dt+\right.\\
\left. \int_0^1\left[(3a_n+U_\epsilon(a_n))\dot
a_n^2-a_n^3\dot\phi_n^2\right]\,\mathrm dt\int_0^1 6a_n^2\alpha
V_\lambda(\phi_n)\,\mathrm
dt\right\vert\le\epsilon_n\Vert\alpha\Vert_{H^1_0}.
\end{multline*}
Taking $\alpha:=a_n-a$, and recalling that
\begin{itemize}
\item $\int_0^1\dot a_n^2\,\mathrm dt$ is bounded,
\item $\Vert U'_\epsilon(a_n)\Vert_{L^\infty}$ is bounded (since $a_n$
doesn't
approach either $m$ or $M$),
\item $a_n\to a$ uniformly, and $a_n$ is bounded,
\item $\int_0^12 a_n^3 V_\lambda(\phi_n)\,\mathrm dt$ is bounded away from
zero, and
\item $V_\lambda(\phi_n)$ is bounded,
\end{itemize}
we obtain
\[
\int_0^1(3a_n+U_\epsilon(a_n))\dot a_n(\dot a_n-a)\,\mathrm dt\to 0.
\]
But $3 a_n+U_\epsilon(a_n)\to 3a+U_\epsilon(a)$ uniformly, and
$3a+U_\epsilon(a)\ge 3a\ge 3a_1>0,\,\forall t\in[0,1]$. Then, arguing as
before, we also obtain \eqref{eq:5.10}, and the proof is complete
\end{proof}

\begin{proof}[Proof of Proposition \ref{thm:prop5.5}]

As already outlined, the aim is to apply Theorem \ref{thm:teo3.4} to the
functional $F_{\epsilon,\lambda}$. In view of lemmas \ref{thm:lem5.3} and
\ref{thm:lem5.4}, it still remains to show that hypothesis \eqref{itm:assi}
holds.

Take $a_*(t)=(1-t)a_0+t a_1$, and choose $\omega_0=a_*$, $e_0=\phi_*$. Note
that, by \eqref{eq:4.5} and the choice of $m$, we have
\begin{equation}\label{eq:5.1}
\int_0^1a^3 V_\lambda(\phi_*)\,\mathrm dt\ge m^3 v_*>0,\qquad\forall
a\in\Omega,\,\forall\lambda\ge\sup_{t\in[0,1]}V(\phi_*).
\end{equation}
Moreover by assumption \eqref{eq:4.2} we have, for any $a\in\Omega$,
\begin{multline*}
\int_0^1\left[\left(3a+U_\epsilon(a)\right)\dot
a^2-a^3\dot\phi_*^2\right]\,\mathrm dt\ge\\
\int_0^1 3m\dot a^2-M^3(\phi_1-\phi_0)^2\,\mathrm dt\ge
3m(a_1-a_0)^2-M^3(\phi_1-\phi_0)^2>0,
\end{multline*}
therefore, by \eqref{eq:5.1},
\begin{equation}\label{eq:5.2}
b_1:=\inf F_{\epsilon,\lambda}(a,\phi_*)>0.
\end{equation}
Clearly, $b_1=b_1(m,M)$ and is independent of $k$. Actually, $b_1$ is
independent of $\lambda$ too, but this property won't be used here.

Since $V_\lambda$ is bounded,
\[
\sup_{\R}V_\lambda\equiv B_\lambda<+\infty,
\]
and
\[
\int_0^1 a_*^3 V_\lambda(\phi)\,\mathrm dt\le a_0^3 B_\lambda.
\]
Moreover, if $\epsilon$ is sufficiently small,
$U_\epsilon(a_*(t))=0\,\forall
t$, so
\[
F_{\epsilon,\lambda}(a_*,\phi)=\int_0^1\left(3 a_*\dot
a_*^2-a_*^3\dot\phi^2\right)\,\mathrm dt\cdot\int_0^12 a_*^3
V_\lambda(\phi)\,\mathrm dt,
\]
but $\int_0^1\left(3 a_*\dot a_*^2-a_*^3\dot\phi^2\right)\,\mathrm
dt\le\int_0^1\left(3a_*\dot a_*^2-m^3\dot\phi^2\right)\,\mathrm dt$, and
then
there exists $R>0$ independent of $k$ such that
\begin{equation}\label{eq:5.3}
\sup_{\Vert\widehat{\phi}\Vert_{E_k}=R}F_{\epsilon,\lambda}(a_*,\widehat{\phi}+\phi_*)=:
b_0<b_1,
\end{equation}
(note that $b_0$ depends on $\lambda$) and then hypothesis \eqref{itm:assi}
of
Theorem \ref{thm:teo3.4} also holds. This implies that there exists a
critical
value for the functional $F_{\epsilon,\lambda}$ on $\mathfrak{X}_k$ in the
interval $[b_1,b_2]$, where
\begin{equation}\label{eq:5.4}
b_2:=\sup_{\Vert\widehat{\phi}\Vert_{E_k}\le
R}F_{\epsilon,\lambda}(a_*,\widehat{\phi}+\phi_*)
\end{equation}
and since $b_2$, though depending on $\lambda$, is independent of $k$, the
proof is complete.
\end{proof}

\section{Proof of the main result}\label{sec:proof}
To complete the proof of Theorem \ref{thm:main}, using Proposition
\ref{thm:prop5.5}, we first need the following lemma. Recall the definitions
of
$b_1$, $b_2$ given in \eqref{eq:5.2} and \eqref{eq:5.4}.

\begin{lem}\label{thm:lem6.1}
There exists a critical value of $F_{\epsilon,\lambda}$ on $\mathfrak
X=\Omega\times E$ in $[b_1,b_2]$.
\end{lem}

\begin{proof}
Let $x_k=(a_k,\phi_k)$ the critical point given by Proposition
\ref{thm:prop5.5}. Arguing as in the proof of Lemma \ref{thm:lem5.4} we
deduce
the existence of $x=(a,\phi)\in\mathfrak{X}$ such that, up to subsequence,
\[
x_k\to x\qquad\text{in\ }H^1.
\]
We will show that $x$ is a critical point of $F_{\epsilon,\lambda}$. Since
$F_{\epsilon,\lambda}(x_k)\in[b_1,b_2]$ and $F_{\epsilon,\lambda}$ is
continuous, it immediately will follow that
$F_{\epsilon,\lambda}(x)\in[b_1,b_2]$.

Take $(\alpha,\theta)$ in $H^1_0$, and consider $\theta_k$, the orthogonal
projection of $\theta$ on $W_k$. As observed in remark \ref{rem:4.1},
$\theta_k\to\theta$ in $H^1$. Since $\nabla
F_{\epsilon,\lambda}(x_k)[\alpha,\theta_k]=0$ for any $k$, we have
\[
\nabla F_{\epsilon,\lambda}(x)[\alpha,\theta]=\nabla
F_{\epsilon,\lambda}(x)[\alpha, \theta-\theta_k]+\left(\nabla
F_{\epsilon,\lambda}(x)-\nabla
F_{\epsilon,\lambda}(x_k)\right)[\alpha,\theta_k].
\]
Let us observe that, since $[\alpha,\theta_k]$ is bounded in $H^1$, and --
recalling that $F_{\epsilon,\lambda}$ is $\mathcal{C}^1$ -- $\nabla
F_{\epsilon,\lambda}(x_k)\to \nabla F_{\epsilon,\lambda}(x)$, then
$\left(\nabla F_{\epsilon,\lambda}(x)-\nabla
F_{\epsilon,\lambda}(x_k)\right)[\alpha,\theta_k]\to 0$. Moreover
\[
\nabla F_{\epsilon,\lambda}(x)[\alpha,\theta-\theta_k]\to 0
\]
since $\theta_k\to\theta$ in $H^1$. Then $\nabla
F_{\epsilon,\lambda}(x)[\alpha,\theta]=0,\,\forall\alpha,\theta$ in $H^1_0$,
and the proof is complete.
\end{proof}

\begin{proof}[Proof of Theorem \ref{thm:main}]
Recall first that, looking for solutions of \eqref{eq:efe1}--\eqref{eq:efe2}
in
the space of curves $(a,\phi)$ defined on $[0,T]$ and with boundary
conditions
\eqref{eq:boundary}, amounts to find critical points for the functional
\eqref{eq:func2} in the space of curves $(a,\phi)$ defined in $[0,1]$ with
boundary conditions \eqref{eq:bound}.

The above lemma ensures the existence of
$(a_{\epsilon,\lambda},\phi_{\epsilon,\lambda})$, critical point of
$F_{\epsilon,\lambda}$ in $\Omega\times E$, with critical value in
$[b_1,b_2]$.
First, let us observe that a simple bootstrap argument shows that both
$a_{\epsilon,\lambda}$ and $\phi_{\epsilon,\lambda}$ are $\mathcal{C}^2$.
Then
the following conservation law follows from the variational principle:
\begin{equation}\label{eq:6.1}
(3a_{\epsilon,\lambda}+U_\epsilon(a_{\epsilon,\lambda})) \dot
a_{\epsilon,\lambda}^2-a_{\epsilon,\lambda}^3\dot\phi_{\epsilon,\lambda}^2 -2
a_{\epsilon,\lambda}^3V_\lambda(\phi_{\epsilon,\lambda})=0.
\end{equation}
Since $a_{\epsilon,\lambda}^3(\dot\phi_{\epsilon,\lambda}^2+
2V_\lambda(\phi_{\epsilon,\lambda}))>0,\,\forall t$, then $\dot
a_{\epsilon,\lambda}(t)\ne 0,\,\forall t$, and since we have supposed (see
\eqref{eq:4.1}) $a_0>a_1$ it is
\[
\dot a_{\epsilon,\lambda}(t)<0,\qquad\forall t\in[0,1]
\]
and
\[
a_1\le a_{\epsilon,\lambda}(t)\le a_0,\qquad\forall t\in[0,1].
\]
Then $a_{\epsilon,\lambda}(t)$ is bounded away from $m$ and $M$. Taking a
sufficiently small $\epsilon$ we have
\[
m+\epsilon<a_{\epsilon,\lambda}(t)<M-\epsilon
\]
so that $U_\epsilon(a_{\epsilon,\lambda})=0$,
$U'_\epsilon(a_{\epsilon,\lambda})=0$ and therefore the couple
$(a_\lambda,\phi_\lambda):=(a_{\epsilon,\lambda},\phi_{\epsilon,\lambda})$
is a
critical point of
\[
F_\lambda(a,\phi):=\int_0^1\left(3a\dot a^2-a^3\dot\phi^2\right)\,\mathrm
dt\cdot\int_0^1 2 a^3 V_\lambda(\phi)\,\mathrm dt,
\]
with critical value in $[b_1,b_2]$. Moreover, the conservation law
\eqref{eq:6.1} takes the form
\begin{equation}\label{eq:6.2}
3 a_\lambda\dot a_\lambda^2-a_\lambda^3\dot\phi_\lambda^2-2 a_\lambda^3
V_\lambda(\phi_\lambda)=0.
\end{equation}
Recalling that $\dot a_\lambda$ is negative, that implies
\[
\frac{\dot
a_\lambda}{a_\lambda}=-\sqrt{\frac{\dot\phi_\lambda^2+2V_\lambda(\phi_\lambda)}3}.
\]
Integrating the above relation in $[0,1]$ we obtain
\[
a_\lambda(t)=a_0
e^{-\frac1{\sqrt3}\int_0^t\sqrt{\dot\phi_\lambda^2+2V_\lambda(\phi_\lambda)}\,\mathrm
ds},
\]
and in particular
\[
a_1=a_0
e^{-\frac1{\sqrt3}\int_0^1\sqrt{\dot\phi_\lambda^2+2V_\lambda(\phi_\lambda)}\,\mathrm
ds}.
\]
But $a_1>0$ is of course fixed and therefore independent of $\lambda$, then
there exists a positive constant $D$ independent of $\lambda$ such that
\[
\int_0^1\vert\dot\phi_\lambda\vert\,\mathrm
dt\le\int_0^1\sqrt{\dot\phi_\lambda^2+2V_\lambda(\phi_\lambda)}\,\mathrm
ds\le
D.
\]
Since $\phi_0$ is fixed, we see that, for any $\lambda$, the function
$\phi_\lambda$
satisfies
\[
\vert\phi_\lambda(t)\vert\le\vert\phi_0\vert+D,
\]
and therefore choosing $\lambda\ge
\sup\{V(\phi)\,:\,\vert\phi\vert\le\vert\phi_0+D\}$ we have
\[
V(\phi_\lambda)=V_\lambda(\phi_\lambda),\qquad
V'(\phi_\lambda)=V'_\lambda(\phi_\lambda).
\]
This means that $(a_\lambda,\phi_\lambda)$ is also a critical point of $F$
\eqref{eq:func2} with positive critical value.
\end{proof}


\end{document}